\documentclass[twocolumn,showpacs,preprintnumbers,amsmath,amssymb,prb,aps]{revtex4}

\usepackage{epsfig}
\usepackage{graphicx}
\usepackage{dcolumn}
\usepackage{bm}

\begin{document}

\title{Role of covalency in the ground state properties of perovskite ruthenates:
A first principle study using local spin density approximations}

\author{Kalobaran Maiti}

\affiliation{Department of Condensed Matter Physics and Material
Sciences, Tata Institute of Fundamental Research, Homi Bhabha
Road, Colaba, Mumbai - 400 005, INDIA}

\date{\today}

\begin{abstract}

We investigate the electronic structure of SrRuO$_3$ and CaRuO$_3$
using full potential linearized augmented plane wave method within
the local spin density approximations. The ferromagnetic ground
state in SrRuO$_3$ could exactly be described in these
calculations and the calculated spin magnetic moment is found to
be close to the experimentally observed values. Interestingly, the
spin polarized calculations for CaRuO$_3$ exhibit large spin
moment as observed in the experiments but the magnetic ground
state has higher energy than that in the non-magnetic solution.
Various calculations for different structural configurations
indicate that Ca-O covalency plays the key role in determining the
electronic structure and thereby the magnetic ground state in this
system.

\end{abstract}


\maketitle

\section{Introduction}

Investigation of the ground state properties of ruthenates has
seen an explosive growth in the recent time due to many
interesting properties such as superconductivity,\cite{sr2ruo4}
non-Fermi liquid behavior,\cite{nfl,klein} unusual magnetic ground
states\cite{nfl,klein,rss,cao} {\it etc.} observed in these
materials. SrRuO$_3$, a perovskite compound exhibits ferromagnetic
long range order below the Curie Temperature of about 160~K with a
large magnetic moment (1.4~$\mu_B$) despite highly extended 4$d$
character of the valence electrons.\cite{rss,cao} Various band
structure calculations based on different kinds of approximations
reveal ferromagnetic ground state in SrRuO$_3$ and the calculated
magnetic moments are found to strongly depend on the kind of
approximations used in the study.\cite{david,vidya} Interestingly,
CaRuO$_3$, an isostructural compound exhibit similar magnetic
moment at high temperatures as that observed in SrRuO$_3$ but the
magnetic ground state is controversial. While some studies
predicted an antiferromagnetic ordering in
CaRuO$_3$,\cite{vidya,callaghan,longo,sugiyama} various other
studies suggest absence of long-range order down to the lowest
temperature studied.\cite{nfl,klein,cao,martinez} These later
investigations predict that the behavior in this compound is in
the proximity of quantum criticality, which is manifested in the
transport measurements exhibiting non-Fermi liquid
behavior.\cite{nfl,klein}

\begin{figure}
\vspace{-4ex}
 \centerline{\epsfysize=4.0in \epsffile{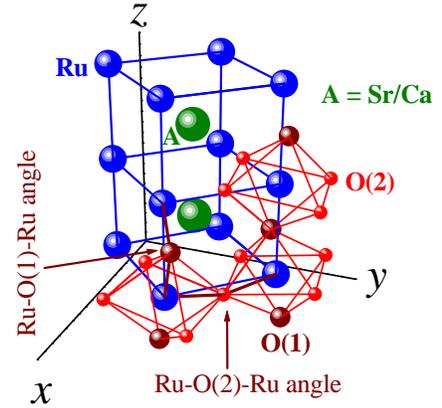}}
\vspace{-20ex}
 \caption{Schematic diagram of the crystal structure of
SrRuO$_3$ and CaRuO$_3$. The thick solid lines show the Ru-O-Ru
angles.}
 \vspace{-2ex}
\end{figure}

Both SrRuO$_3$ and CaRuO$_3$ form in an orthorhombic perovskite
structure (ABO$_3$ - type). The space group for SrRuO$_3$ is
conventionally defined as $Pbnm$ and that for CaRuO$_3$ is $Pnma$,
which are essentially the same structure type with a difference in
the definition of axis system.\cite{ramarao,nakatsugawa} The A
cation (Sr/Ca) sites help to form the typical building block of
this structure. The RuO$_6$ octahedra in these compounds are
connected by corner sharing as shown in Fig.~1. The conduction
electrons moves via this RuO$_6$ network and hence determines
various electronic properties. The tilting and buckling of the
RuO$_6$ octahedra as evident in the figure essentially leads to a
GdFeO$_3$ type distortion resulting to the orthorhombic structure.
While the structure type is same in both the compounds, the extent
of distortion is somewhat different resulting to a slightly
different Ru-O-Ru bond angle in these compounds. If O(1)
represents the apical oxygen in the RuO$_6$ octahedra along
$z$-axis in the figure and O(2) represents the oxygens in the
basal ($xy$) plane, there are two O(2) and one O(1) atoms in one
formula unit. The Ru-O-Ru angles in SrRuO$_3$
are:\cite{nakatsugawa} Ru-O(1)-Ru = 167.6$^\circ$, Ru-O(2)-Ru =
159.7$^\circ$ and those in CaRuO$_3$ are:\cite{kobayashi}
Ru-O(1)-Ru = 149.6$^\circ$, Ru-O(2)-Ru = 149.8$^\circ$.

The Ru-O-Ru angle has significant influence in the electronic
properties since the electron hopping interaction strength between
Ru-sites via oxygens, usually denoted by $t$ is the largest for
the angle of 180$^\circ$ and gradually reduces with the deviation
of Ru-O-Ru bond angle from 180$^\circ$. Thus, the effective
electron correlation strength, $U/W$ ($U$ = electron-electron
Coulomb repulsion strength and $t \propto W$ = valence band width)
is expected to be higher in CaRuO$_3$ than that in SrRuO$_3$. It
was believed that the difference in $U/W$ in these two compounds
leads to such contrasting ground state properties. However, a
recent experimental study shows that $U/W$ is significantly weak
as expected for a highly extended 4$d$ transition metal oxides and
are very similar in {\it both} the compounds.\cite{ravi} Thus, the
experimental observation of different ground state properties in
otherwise these two identical compounds still remains a puzzle
despite numerous studies carried out in these and associated other
systems as well during last two decades.

Local spin density approximations has often been found to be
successful to capture the magnetic ground state properties of
various systems.\cite{dd,hamada,kbm} In the present study, we
investigate the magnetic ground state properties of ruthenates
using {\em state-of-the-art} full potential linearized augmented
plane wave method\cite{wien} within the local spin density
approximations. The results for SrRuO$_3$ reveal ferromagnetic
ground state consistent with various experimental results.
Interestingly, the ground state energy for the spin polarized
solution of CaRuO$_3$ is higher than that for the non-magnetic
solution despite large exchange splitting leading to a large
magnetic moment in this system. Calculations for different
structural configurations suggest that Ca-O covalency plays the
key role in determining the electronic properties in this system.

\section{Theoretical methods}

The electronic band structure calculations were carried out using
full potential linearized augmented plane wave (FLAPW) method
within the local spin density approximations (LSDA) using
{\scriptsize WIEN2K} software.\cite{wien} The lattice constants
are obtained from the analysis of $x$-ray diffraction patterns
from high quality polycrystalline samples\cite{rss} and are very
similar to those estimated earlier for the samples in {\it both}
single crystalline and polycrystalline
forms.\cite{ramarao,nakatsugawa,kobayashi} The estimated lattice
parameters corresponding to the axis system shown in Fig.~1 are;
$a$ = 5.572~\AA, $b$ = 5.542~\AA, $c$ = 7.834~\AA\ for SrRuO$_3$,
and $a$ = 5.519~\AA, $b$ = 7.665~\AA, $c$ = 5.364~\AA\ for
CaRuO$_3$. The atomic positions in SrRuO$_3$ used in these
calculations are; Sr: 4$c$ ($x$=0.488(3), $y$=1.00(0), 1/4); Ru:
4$b$ (0, 0.5, 0); O(1): 4$c$ ($x$=-0.000(1), $y$=0.03(8), 1/4);
O(2): 8$d$ ($x$=0.28(5), $y$=0.22(7), $z$=-0.03(3)). The atomic
positions in CaRuO$_3$ are; Ca: 4$c$ ($x$=0.9465(7), 1/4,
$z$=0.0122(5)); Ru: 4$b$ (0, 0, 0.5); O(1): 4$c$ ($x$=0.10248(7),
1/4, $z$=0.5899(1)) and O(2): 8$d$ ($x$=0.2072(5), $y$=0.4513(5),
$z$=0.2080(8)). It is to note here that the Ru 4$d$ orbitals
defined in the axis system shown in Fig.~1 are different from
those in the local axis system of RuO$_6$ octahedra having
$t_{2g}$ and $e_g$ symmetries. In order to show the partial
density of states corresponding to different Ru 4$d$ orbitals, we
have rotated the axis system by 45$^\circ$ in the $xy$ plane and
the results are shown later in the text.

Various bond lengths found in these calculations match well with
the experimental results. Despite difference in lattice parameters
and change in unit cell volume between CaRuO$_3$ and SrRuO$_3$,
average Ru-O bond lengths are very similar in {\it both} the
cases. The muffin-tin radii ($R_{MT}$) for Sr/Ca, Ru and O were
set to 1.06~\AA, 0.85~\AA\ and 0.74~\AA, respectively. The
convergence for different calculations were achieved considering
512 $k$ points within the first Brillouin zone. The error bar for
the energy convergence was set to $<$~0.25~meV per formula unit.
In every case, the charge convergence was achieved to be less than
10$^{-3}$ electronic charge.

\section{Results and Discussions}

\begin{figure}
\vspace{-4ex}
 \centerline{\epsfysize=4.5in \epsffile{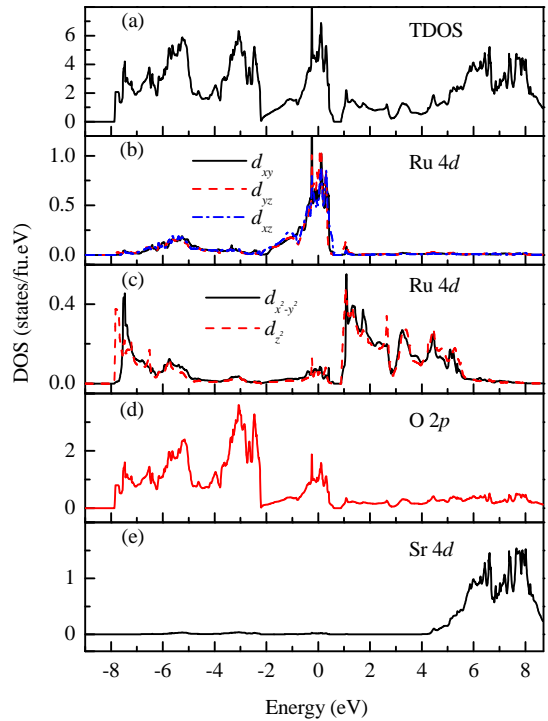}}
\vspace{-4ex}
 \caption{Calculated (a) TDOS, (b) Ru 4$d$ PDOS with
$t_{2g}$ symmetry, (c) Ru 4$d$ PDOS with $e_g$ symmetry, (d) O
2$p$ PDOS, and (e) Sr 4$d$ PDOS for the non-magnetic ground state
of SrRuO$_3$. The large intensity at the Fermi level suggests a
metallic ground state. The states at the Fermi level have
primarily Ru 4$d$ character with O 2$p$ states appearing below
-~2~eV.}
 \vspace{-2ex}
\end{figure}

In Fig.~2, we show the density of states (DOS) calculated for the
non-magnetic solution of SrRuO$_3$. There are four distinct groups
of intense features and a flat region appear in the total density
of states (TDOS) shown in Fig.~2(a). In order to identify the
character of these features, we also plot the partial density of
states (PDOS) corresponding to Ru~4$d$, O~2$p$ and Sr~4$d$
contributions in the eigen states by calculating the projection of
the eigen states onto these spin-orbitals. It is clear that
Sr~4$d$ contributions appear beyond 4~eV above the Fermi level,
$\epsilon_F$ denoted by 'zero' in the energy axis. O~2$p$ also has
finite contributions in this energy range suggesting finite mixing
between O~2$p$ and Sr~4$d$ electronic states.

The Ru~4$d$ contributions are shown in two groups. In Fig.~2(b)
and 2(c), the 4$d$ orbitals defined in the cartesian coordinate
systems correspond to the $xyz$-axis system rotated by 45$^\circ$
in the $xy$ plane with respect to the axis system shown in Fig.~1.
This axis system is very close to the local axis system of the
RuO$_6$ octahedra. The deviations due to the orthorhombic
distortions are clearly manifested by small intensities in the
energy range corresponding to other symmetries. The $d_{xy}$,
$d_{yz}$ and $d_{xz}$ orbitals possess $t_{2g}$ symmetry and are
shown in Fig.~2(b). The PDOS corresponding to these orbitals
spread over a large energy range of -~7~eV to 0.6~eV with a sharp
feature between -~2~eV to 0.6~eV. O~2$p$ PDOS in Fig~2(d) also
exhibit significantly large intensities in this energy range. The
observation of O~2$p$ character in the Ru~4$d$ band and vice versa
suggests strong hybridization between O~2$p$ and Ru~4$d$
electronic states. This hybridization between $t_{2g}$ bands and
the symmetry adapted O 2$p$ bands forms bonding and anti-bonding
energy bands, where the overlap of the orbitals are sideways and
are known as $\pi$-bonds. The energy bands between -~7~eV to
-~4.5~eV energies can be assigned as bonding bands and the
anti-bonding bands appear between -~2 to 0.6~eV. From the relative
intensities of the Ru 4$d$ and O 2$p$ PDOS, it is evident that the
anti-bonding bands possess essentially Ru 4$d$ character and the
bonding bands have O 2$p$ character. The total width of the
$t_{2g}$ band is close to 2.6~eV.

\begin{figure}
\vspace{-4ex}
 \centerline{\epsfysize=4.5in \epsffile{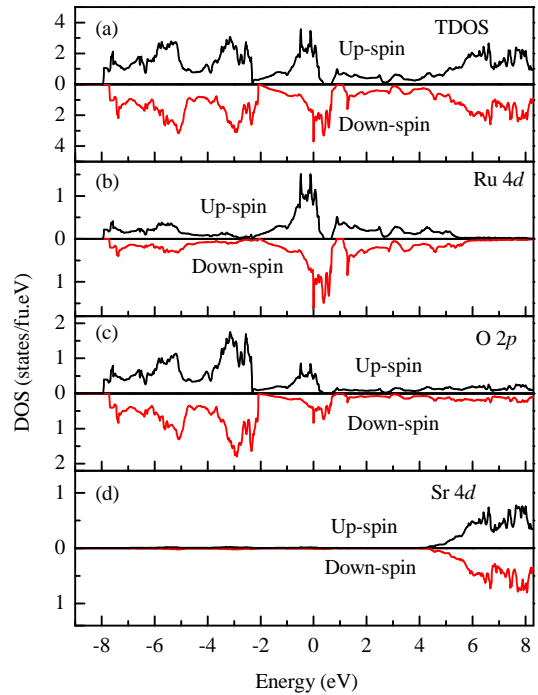}}
\vspace{-8ex}
 \caption{Calculated (a) TDOS, (b) Ru 4$d$ PDOS, (c) O
2$p$ PDOS, and (d) Sr 4$d$ PDOS for the ferromagnetic ground state
of SrRuO$_3$. The down spin DOS are shown in the same energy scale
with inverted DOS axis.}
 \vspace{-2ex}
\end{figure}

The $d_{x^2-y^2}$ and $d_{z^2}$ PDOS are shown in Fig.~2(c). The
primary contribution appears between 0.9~eV and 5.5~eV with large
contributions between -~7.8~eV to -~5~eV. Similar to the case of
$t_{2g}$ bands, significant O 2$p$ PDOS also appear in these
energy ranges. The mutual contributions between Ru 4$d$ and O 2$p$
states again indicate large degree of covalency. The electronic
states with $e_g$ symmetry form $\sigma$-bonds (head-on overlap)
with the O~2$p$ states. Since, $\sigma$-bonds are significantly
stronger than the $\pi$-bonds, it is expected that the separation
between bonding and anti-bonding states involving $\sigma$-bonds
will be significantly larger compared to that corresponding to
$\pi$-bonds. This is clearly manifested in the electronic density
of states in Fig.~2(c), where the energy region between -~7.8~eV
and -~5~eV is contributed by bonding bands with dominant O 2$p$
character and the anti-bonding states appear above $\epsilon_F$.

The features in the energy region between -~4~eV to -~2.2~eV are
primarily contributed by O 2$p$ PDOS, and has negligible
contributions from Ru~4$d$ and Sr 4$d$ electronic states. Thus,
these intensities are assigned to the O~2$p$ non-bonding
electronic states. Notably, all these characterizations are
consistent with various photoemission results.\cite{ravi}

In order to investigate the magnetic properties in this system
within the local spin-density approximations, we have calculated
the ground state energies for ferromagnetic arrangement of moments
of the constituents. Interestingly, the eigen energy for the
ferromagnetic ground state is lower by 1.2~meV per formula unit
(fu) than the lowest eigen energy for the non-magnetic solution.
It is to note here that a small variation in atomic position leads
to a large difference in energy between the ground state energies
corresponding to ferromagnetic and non-magnetic solutions. For
example, for the atomic positions from Ref.~[12], this energy
difference is about 9~meV~/fu. However, the ferromagnetic ground
state energy in this case is 612~meV~/fu higher than that in the
present case. Interestingly, all the structures of SrRuO$_3$
published in the literature to my knowledge, the ferromagnetic
ground state energy is observed to be lower than that
corresponding to the non-magnetic solution. While these results
are clearly consistent with the experimental observations and
various previous results,\cite{david,vidya} it reestablishes that
the local spin density approximations are sufficient to capture
the magnetic ground state in these materials as also established
in 3$d$ and 5$d$ transition metal oxides.\cite{dd,hamada,kbm}

The spin magnetic moment centered at Ru-sites is found to be about
0.58~$\mu_B$, which is substantially small compared to the spin
only value of 2~$\mu_B$ corresponding to four electrons in the
$t_{2g}$ orbitals in the low spin configuration of Ru$^{4+}$. In
addition, there is a large spin polarization of the interstitial
electronic states ($\sim$~0.33~$\mu_B$ $\Rightarrow$ about 57\% of
the Ru-moment). The induced spin moment at the oxygen sites are
also found to be significantly large (0.1~$\mu_B$ for O(1) and
0.08~$\mu_B$ for O(2)). All these magnetic moments centered at
different Ru and O sites couple ferromagnetically. Thus, the total
spin magnetic moment per formula unit is found to be about
1.2~$\mu_B$ and is very close to the experimental estimation of
1.4~$\pm$~0.4~$\mu_B$ from various measurements on varieties of
samples in single and polycrystalline
forms.\cite{longo,kanbayashi}

Interestingly, the total magnetic moment, although consistent to
the experimental estimations, is significantly lower than the spin
only value for the $t_{2g\uparrow}^3t_{2g\downarrow}^1$ electronic
configuration for Ru$^{4+}$ ions. This is significantly different
from 3$d$ transition metal oxides, where the calculated magnetic
moment is often found very close to their spin only value
corresponding to the electronic configuration close to the
transition metal ions. While it has been observed that various
other approximations, such as generalized gradient approximations,
inclusion of spin-orbit coupling {\it etc.} leads to a larger
value of the spin magnetic moment in the ground state, the
experimental observation of small saturation magnetic moment even
for high quality single crystals suggests that other effects are
important for this system. In particular, the orbital moment is
expected to be negligible for these large 4$d$ orbitals due to the
strong crystal field effect. The reduction of local moment from
the spin only value of 2~$\mu_B$ may be attributed to the highly
extended nature of the 4$d$ orbitals. In addition, a large degree
of O 2$p$ - Ru 4$d$ hybridization is observed in the DOS, which
may reduce the magnetic moment further.

The shifts in the DOS due to exchange splitting are shown in
Fig.~3, where the down-spin DOS is shown on the same energy scale
($x$-axis) but the DOS-axis is inverted. The relative shift of the
up- and down-spin states provides an estimation of the exchange
splitting of the corresponding electronic states. It is evident
that there is no shift between up- and down-spin Sr~4$d$ states
appearing above 4~eV. The relative shift in the up- and down-spin
PDOS corresponding to O 2$p$ states provides an estimate of the
exchange splitting of about 0.25~eV. The exchange splitting in Ru
4$d$ total density of states is close to 0.5~eV and is consistent
with the values obtained in previous studies.\cite{david} However,
it is found to be somewhat different for the electronic states
with different symmetries. While it is 0.5~eV for $t_{2g}$ bands,
close to 0.4~eV is observed in the case of $e_g$ bands. Smaller
splitting in the later cases may be attributed to the larger
degree of itineracy of the corresponding electronic states.
Interestingly, the TDOS at $\epsilon_F$ for the up-spin states is
found to be significantly smaller (1.95 states/eV.fu) than that
(3.64 states/eV.fu) for the down spin states. The spin
polarization can be defined as $$P = {{(N(\epsilon_F)_\uparrow -
N(\epsilon_F)_\downarrow)} \over {(N(\epsilon_F)_\uparrow +
N(\epsilon_F)_\downarrow)}}$$ where $N(\epsilon_F)$ denotes the
density of state at $\epsilon_F$. Thus, $P$ is found to be
-~30.2\%, which is significantly large and negative as observed
experimentally.\cite{worledge}

\begin{figure}
\vspace{-4ex}
 \centerline{\epsfysize=4.5in \epsffile{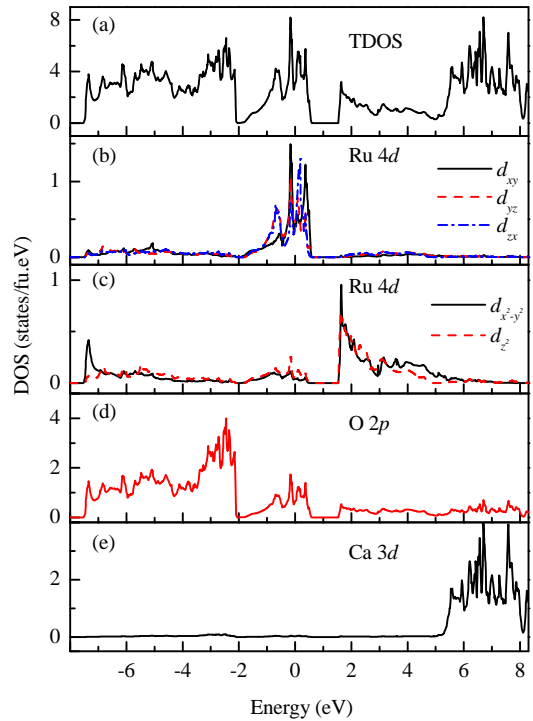}}
\vspace{-4ex}
 \caption{Calculated (a) TDOS, (b) Ru 4$d$ PDOS with
$t_{2g}$ symmetry, (c) Ru 4$d$ PDOS with $e_g$ symmetry, (d) O
2$p$ PDOS, and (e) Ca 3$d$ PDOS for the non-magnetic ground state
of CaRuO$_3$. The large intensity at the Fermi level suggests a
metallic ground state as that observed in SrRuO$_3$. The states at
the Fermi level have primarily Ru 4$d$ character with O 2$p$
states appearing below -~2~eV.}
 \vspace{-2ex}
\end{figure}

We now turn to the case of CaRuO$_3$. The calculated DOS for
non-magnetic solutions are shown in Fig.~4. The features below
-~2~eV is primarily contributed by the O~2$p$ PDOS as also
observed in the case of SrRuO$_3$ in Fig.~2. The oxygen 2$p$
non-bonding bands appear between -~4~eV to -~2~eV energy range. By
comparing the Ru 4$d$ PDOS and O 2$p$ PDOS, it is evident that the
bonding $t_{2g}$ and $e_g$ bands with dominant O 2$p$ character
appears below -~4~eV, where the bonding $e_g$ bands have the lower
energies as expected. The anti-bonding $t_{2g}$ bands possess
dominant Ru 4$d$ character and appear between -~1.8~eV to 0.6~eV
energies. The anti-bonding $e_g$ bands appear between the energy
range of 1.5~eV to 5~eV. The $t_{2g}$ and $e_g$ bands are
separated by a distinct energy gap of about 1~eV. Interestingly,
{\it Ca 3$d$ contributions appear above 5~eV, which is
significantly higher than the energy range corresponding to the Sr
4$d$ contributions.} This is unusual since the Sr 4$d$ electronic
states are expected to have higher energies compared to the
energies of Ca 3$d$ electronic states. We will discuss this
unusual observation later in the text.

\begin{figure}
\vspace{-4ex}
 \centerline{\epsfysize=4.5in \epsffile{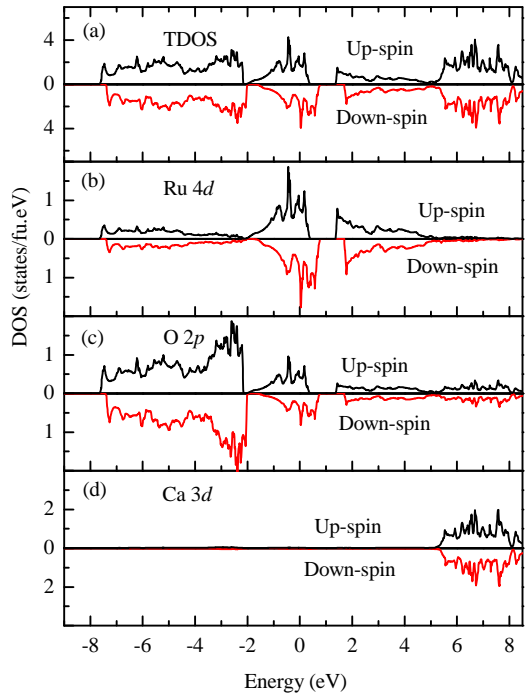}}
\vspace{-8ex}
 \caption{Calculated (a) TDOS, (b) Ru 4$d$ PDOS, (c) O
2$p$ PDOS, and (d) Ca 3$d$ PDOS for the ferromagnetic ground state
of CaRuO$_3$. The down spin DOS are shown in the same energy scale
with inverted DOS axis.}
 \vspace{-2ex}
\end{figure}

It is to note here that the total width of the $t_{2g}$ band is
about 2.4~eV, which is slightly smaller than that observed for
SrRuO$_3$ ($\sim$~2.6~eV; a change of 7.5\%). However, the width
of the individual $d$-bands are very close to each other
($\sim$~2.4~$\pm$~0.1~eV). Thus, the smaller Ru-O-Ru bond angle in
CaRuO$_3$ has very little influence on the width of the $t_{2g}$
bands. While all the $d$ bands with $t_{2g}$ symmetry are very
similar in SrRuO$_3$, the $d_{xy}$ band in CaRuO$_3$ is slightly
shifted to higher energies compared to the $d_{yz}$ and $d_{xz}$
bands. This energy shift may be attributed to the distortion of
the RuO$_6$ octahedra leading to a deviation from octahedral
symmetry towards $D_{4h}$ symmetry. The width of the $e_g$ band is
found to be smaller ($\sim$~3.5~eV) in CaRuO$_3$ than that
($\sim$~4.6~eV) in SrRuO$_3$. Since, $e_g$ states are coupled with
oxygens via $\sigma$ bonds, the change in Ru-O-Ru bond angle has
the strongest influence in these bands in comparison to that in
the $\pi$-bonded case corresponding to $t_{2g}$ bands. However,
such changes will not have significant influence in the electronic
properties of these materials since the $e_g$ bands are
essentially empty in both the compounds.

In order to probe the magnetic properties in this system, we
calculated the lowest energy eigen states for the ferromagnetic
ordering of the spin moments at various sites. Interestingly, the
calculations converges to a ferromagnetic ground state with large
spin moments at different sites and the lowest energy eigenvalue
is slightly (4~meV~/fu) higher than that for the non-magnetic
solution. A recent study\cite{vidya} using full potential
linearized muffin-tin orbital (FPLMTO) calculations within
generalized gradient approximations (GGA) and spin-orbit coupling
shows that CaRuO$_3$ exhibit $G$-type antiferromagnetic long range
order, where all the neighboring Ru-sites are
antiferromagnetically coupled. However, in the absence of
spin-orbit coupling, the ground state is found to be
ferromagnetic. There are strong controversy in the ground state
behavior in CaRuO$_3$. The recent studies suggest an enhanced
paramagnetic ground state\cite{cao} and/or proximity to the
quantum criticality.\cite{nfl} Slightly higher energy
corresponding to the ferromagnetic ground state compared to the
non-magnetic/paramagnetic solutions in this study suggests that
ferromagnetic ground state in not stable in this systems as also
observed in the experimental results.

In the ferromagnetic solution, the spin magnetic moment centered
at the muffin-tin spheres corresponding to Ru, O(1), O(2) and
interstitial states are found to be 0.49~$\mu_B$, 0.04~$\mu_B$,
0.08~$\mu_B$ and 0.25~$\mu_B$, respectively. All the spin magnetic
moments at different sites couple ferromagnetically, which leads
to a total spin magnetic moment of about 1~$\mu_B$. This value is
again significantly smaller than the spin only value of 2~$\mu_B$
for Ru$^{4+}$ but consistent with the experimentally observed
values.\cite{rss,cao} In order to investigate the up- and down
spin DOS corresponding to various elements in CaRuO$_3$, we show
the calculated DOS in Fig.~5 in the same way as shown in Fig.~3
for SrRuO$_3$. The exchange splitting in O 2$p$ PDOS is found to
be about 0.25~eV. The exchange splitting in {\it both} $t_{2g}$
and $e_g$ bands is close to 0.4~eV, which is slightly smaller than
that observed in SrRuO$_3$. No exchange splitting is observed for
Ca 3$d$ states. The TDOS at $\epsilon_F$ for the up and down spin
states are 1.75 states/eV.fu and 2.22 states/eV.fu, respectively.
Thus, the spin polarization at $\epsilon_F$ in this compound is
found to be about -~11.9\%.

From the results presented so far, it is clear that the
observation of ferromagnetism in SrRuO$_3$ can be explained by
band structure effects. The lowest eigen energies for different
magnetic arrangements in CaRuO$_3$ suggest a
nonmagnetic/paramagnetic ground state. In order to investigate the
origin of such difference in CaRuO$_3$ compared to SrRuO$_3$, we
calculated the lowest energy eigen states for different
combinations of crystal structures and their magnetic ground
states. As a first step, we calculated the lowest energy solutions
for SrRuO$_3$ in the same crystal structure as that of CaRuO$_3$.
Interestingly, the lowest eigen energy for the nonmagnetic
solution is 1117.5~meV~/fu higher than the ferromagnetic ground
state energy in its real structure. The lowest eigen energy for
the ferromagnetic solutions in this crystal structure is even
higher (1118.5~meV~/fu). The total spin magnetic moment is found
to be very very small (0.14~$\mu_B$). Change in structure reduces
the unit cell volume by about 6.4\% and the average Ru-O-Ru bond
angle from 165$^\circ$ to 150$^\circ$. However, average Ru-O bond
lengths are very similar. The large increase in energy indicates
that such changes does not lead to a stable state. Interestingly,
in the CaRuO$_3$ structure, the ferromagnetic ordering leads to an
enhancement in eigen energies compared to the non-magnetic
solutions although Sr is present at the A-sites.

We now compare the lowest eigen energies calculated for CaRuO$_3$
in the crystal structures similar to that of SrRuO$_3$. The lowest
eigen energy increases by 893.6~meV~/fu compared to the ground
state energy in its original structure in the
nonmagnetic/paramagnetic configuration. Interestingly, the
ferromagnetic solution is 869.7~meV~/fu higher in energy, which is
23.9~meV~/fu lower than that in the nonmagnetic configuration. The
magnetic moment calculated in this configuration is 0.66~$\mu_B$,
0.05~$\mu_B$, 0.12~$\mu_B$ and 0.39~$\mu_B$ for Ru, O(1), O(2) and
interstitial states, respectively. The total magnetic moment is
found to be 1.35~$\mu_B$, which is significantly higher than that
found even in SrRuO$_3$. Thus, in SrRuO$_3$ structure, the
presence of Ca at the A-sites also stabilizes ferromagnetic ground
state. Since the change in structure is essentially reflected in
the Ru-O-Ru bond angle without much change in Ru-O bond length, it
is clear that this change in bond-angle helps to stabilize
ferromagnetic ground state. {\em The A-site potential in the
ABO$_3$ structure has negligible influence in determining the
magnetic ordering of the compound.}

\begin{figure}
\vspace{-4ex}
 \centerline{\epsfysize=4.5in \epsffile{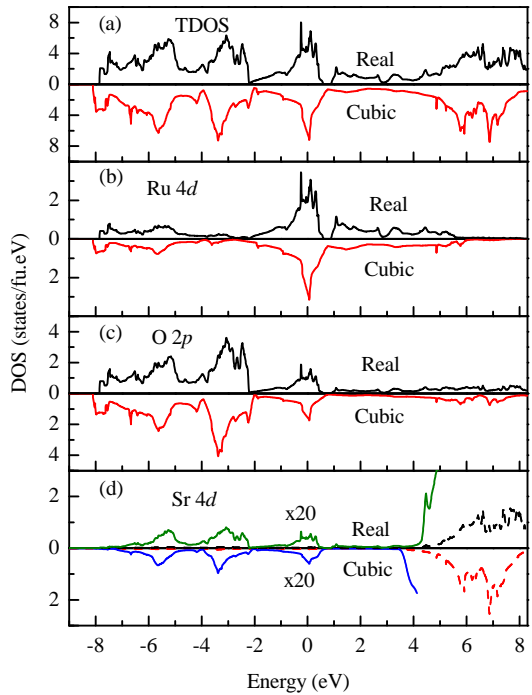}}
\vspace{-8ex}
 \caption{Comparison of (a) TDOS, (b) Ru 4$d$ PDOS, (c)
O 2$p$ PDOS and (d) Sr 4$d$ PDOS of SrRuO$_3$ in the real
structure (positive DOS axis) and equivalent cubic structure
(inverted DOS axis). Results exhibit a close similarity between
the two cases as expected in SrRuO$_3$. In (d), the dashed lines
represent the calculated DOS and the solid lines represented the
same multiplied by 20.}
 \vspace{-2ex}
\end{figure}

In order to probe the role of cations in stabilizing the crystal
structure, which is manifested by large change in the ground state
energies, we compare the calculated results for both SrRuO$_3$ and
CaRuO$_3$ for their real structure and the equivalent cubic
structure keeping the unit cell volume identical to that in the
real structure. The ground state energy for the nonmagnetic
solution of SrRuO$_3$ is about 45~meV~/fu higher than the
ferromagnetic ground state energy in the real structure. The
calculated DOS are shown in Fig.~6. It is evident that the DOS for
real and cubic structures are similar in terms of energy position
and overlap of Ru 4$d$ and Sr 4$d$ electronic states with the O
2$p$ states {\it etc.} Total width of various bands are somewhat
narrower in the real structure compared to that in the cubic
structure. It is to note here that GdFeO$_3$ type of distortion
appears in SrRuO$_3$ via the distortion of the RuO$_6$ octahedra
with different Ru-O bond lengths, and Ru-O(1)-Ru bond angle
167.6$^\circ$ and Ru-O(2)-Ru bond angle 159.7$^\circ$. This leads
to a lifting of degeneracy of the $t_{2g}$ and $e_g$ bands.
However, the average Ru-O bond lengths are close to that in the
equivalent cubic structure. This is clearly manifested by the
closeness of the DOS and PDOS shown in Fig.~6.

\begin{figure}
\vspace{-4ex}
 \centerline{\epsfysize=4.5in \epsffile{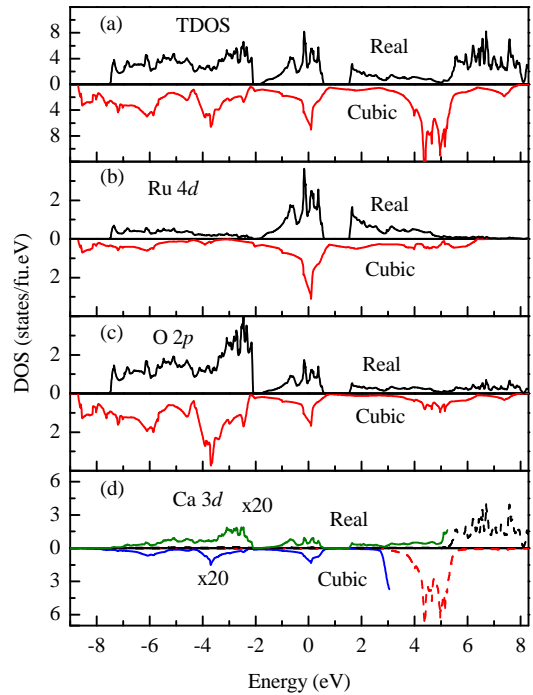}}
\vspace{-8ex}
 \caption{Comparison of (a) TDOS, (b) Ru 4$d$ PDOS, (c)
O 2$p$ PDOS and (d) Ca 3$d$ PDOS of CaRuO$_3$ in the real
structure (positive DOS axis) and equivalent cubic structure
(inverted DOS axis). A large shift of the Ca 3$d$ band suggests
strong Ca-O covalency. In (d), the dashed lines represent the
calculated DOS and the solid lines represented the same multiplied
by 20.}
 \vspace{-2ex}
\end{figure}

In CaRuO$_3$, the nonmagnetic ground state energy in the cubic
structure is significantly large ($\sim$~190~meV~/fu) compared to
that in the real structure. We compare the calculated DOS
corresponding to the real structure of CaRuO$_3$ and the
equivalent cubic structure in Fig.~7. The width of the Ru 4$d$ and
O 2$p$ bands are significantly larger in the cubic structure due
to the Ru-O-Ru bond angle of 180$^\circ$ and a smaller Ru-O bond
length arising from smaller unit cell volume. The most striking
effect is observed in the PDOS of Ca 3$d$ band. In the results
corresponding to the cubic structure, Ca 3$d$ band appear in the
energy range of 3 - 5.5~eV as expected for Ca 3$d$ electronic
states compared to the energies of Sr 4$d$ electronic states shown
in Fig.~6. Such lower binding energies leads to a significant
overlap with the O 2$p$ PDOS appearing at lower energies.
Interestingly, the Ca 3$d$ PDOS contributions shift to above 5~eV
energies in the real structure with an increase in contribution in
the O 2$p$ dominated region as shown by expanding the low energy
part in Fig.~7(d).

\begin{figure}
\vspace{-4ex}
 \centerline{\epsfysize=4.5in \epsffile{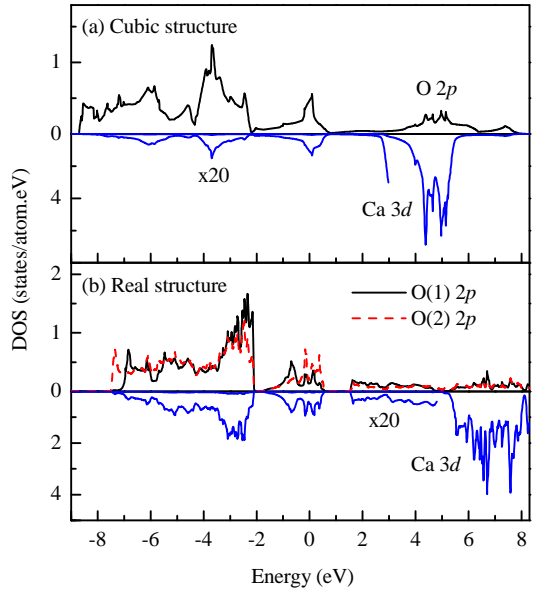}}
\vspace{-18ex}
 \caption{(a) Ca 3$d$ PDOS is compared with O 2$p$ PDOS
in the cubic structure, where the unit cell volume is identical to
the unit cell volume in the real structure. (b) Ca 3$d$ PDOS is
compared with O(1) 2$p$ PDOS and O(2) 2$p$ PDOS corresponding to
the real structure of CaRuO$_3$. The results suggest a strong
Ca-O(1) covalency.}
 \vspace{-2ex}
\end{figure}

In order to investigate the Ca-O covalency and the overlap with
different oxygens in the structure, we plot Ca 3$d$ PDOS along
with corresponding O(1) 2$p$ and O(2) 2$p$ PDOS in Fig.~8. In the
cubic structure, Ca 3$d$ PDOS appears between 3~eV and 5.5~eV
energies. The contributions at lower energies ($E < \epsilon_F$)
appears to have different intensity distribution with energy
compared to that observed for O 2$p$ PDOS. In the real structure,
the Ca 3$d$ band shifts to above 5~eV energies and the
contributions below $\epsilon_F$ matches significantly well with
the PDOS corresponding to O(1) 2$p$ electronic states. This
reveals a typical scenario of Ca 3$d$ and O(1) 2$p$ hybridization
resulting to bonding and antibonding bands. The bands at lower
energies represent the bonding electronic states with dominant O
2$p$ character and the antibonding levels with large Ca 3$d$
character shifts to higher energies. Thus, {\it the DOS in the
real structure reveals a large degree of covalency between Ca-O(1)
in the structure, which leads to a shift of O(1) sites towards
Ca-sites (GdFeO$_3$ type of distortion)}.\cite{andersen}

Smaller ionic radius of Ca$^{2+}$ compared to Sr$^{2+}$ leads to a
reduction in Goldschmidt tolerance factor
$(r_A+r_O)/\sqrt{2}(r_{Ru}+r_O)$, where $r_A$ is the ionic radius
of Sr/Ca. Such reduction often results to a distortion in the
crystal structure. However, this does not indicate, what kind of
distortion is expected. In 3$d^1$ perovskite systems, it is
already shown that although there is no change in the tolerance
factor between CaVO$_3$ and LaTiO$_3$, a large distortion in the
crystal structure is observed.\cite{andersen} It was shown that
the covalency between A-site cations and O-2$p$ states plays the
key role in determining the distortion in the crystal structure
and that the shift of O(1) sites towards A-sites leads to a
GdFeO$_3$-type of distortion in these systems. A similar effect is
also observed here in the electronic structure of these perovskite
ruthenates.

In order to understand the magnetic ground state in CaRuO$_3$ in
the cubic structure, we have calculated the lowest eigen energies
for the same unit cell. Interestingly, in the cubic structure, the
ground state is ferromagnetic, which is about 21~meV~/fu lower in
energy than that for non-magnetic solution. The magnetic moment is
found to be significantly large with total spin contribution of
1.2~$\mu_B$. The Ru-O bond length in this condition is about
1.92~\AA, which is significantly smaller than that in the real
structure (average bond length ~$\cong$~2~\AA). It has been
observed that application of pressure leads to a decrease in Curie
temperature in SrRuO$_3$.\cite{neumeier} This has also been
observed in our calculations as well as in the theoretical results
published in the past\cite{david} that ferromagnetic solution
becomes less stable with the decrease in unit cell volume leading
to a decrease in Ru-O-Ru bond angle. Thus, {\it the observation of
ferromagnetic ground state in the cubic structure of CaRuO$_3$
clearly reveals an important role of Ru-O-Ru bond angle in the
ground state properties.}

In 3$d$ transition metal (TM) oxides, it is observed that the
TM-O-TM superexchange interaction stabilizes antiferromagnetic
ordering for TM-O-TM bond angles of 180$^\circ$. This is the
origin of antiferromagnetic behavior observed in almost all the
3$d$ transition metal oxides with insulating ground state. For
TM-O-TM bond angle of 90$^\circ$, a ferromagnetic ordering is
expected since the oxygen 2$p$ orbitals connected to two
neighboring TM sites are orthogonal and the holes transferred from
TM sites will be parallel due to Hund's rule coupling. Thus, a
deviation from 180$^\circ$ is expected to help ferromagnetic
ordering due to Hund's rule effect.

In addition, the deviation of Ru-O-Ru bond angle from 180$^\circ$
is expected to enhance the local magnetic moment of 4$d$
electronic states due to the decrease in hopping interactions
strength. Thus, {\it the destabilization of the long range
ferromagnetic order in CaRuO$_3$ due to the decrease in Ru-O-Ru
angle from 180$^\circ$ is unusual.}

It is to note here that in 3$d$ transition metal oxides, the
magnetic moment at the transition metal site is often found to be
very close to the values corresponding to the ionic
configuration\cite{dd,hamada} suggesting significantly large
localized character of the 3$d$ electrons. However, 4$d$ orbitals
are highly extended compared to the 3$d$ orbitals and the
correlation effects are found to be significantly weak.\cite{ravi}
This is clearly manifested in various magnetic measurements with
the estimations of significantly smaller magnetic moment (close to
1.4~$\mu_B$) compared to the spin-only value of 2~$\mu_B$ for
Ru$^{4+}$. The extended nature of the 4$d$ electronic states leads
to a strong coupling between the local moments centered at
different cites. This is presumably the origin of large degree of
spin polarization in the O~2$p$ bands ($\geq$~17\%) and in the
interstitial electronic states ($\geq$~55\%). Such large spin
polarization of the valence electronic states has also been
observed in elemental ferromagnets; 4$f$ moments leads to a strong
spin polarization in the 5$d$6$s$ valence electrons of
Gd\cite{gdprl} and these electronic states couple the local
moments at different sites leading to a ferromagnetic long range
order. Such coupling is will be weaker in these ruthenates, when
the Ru-O-Ru angle deviates from 180$^\circ$ due to the reduction
in hopping interaction strength and hence the reduction in
extended nature of the valence electrons. While these arguments
provide a qualitative understanding of the magnetic properties in
these interesting class of materials, further study is required to
probe the details of such interactions in these systems.

In summary, the results in these calculations establishes that A-O
covalency in the ARuO$_3$ structure plays the key role in
determining the electronic structure in these systems. The absence
of long range order in CaRuO$_3$ may be attributed to smaller
Ru-O-Ru angle appearing due to large Ca-O covalency in this
system.

\end{document}